\input{psfig}
\documentstyle[prl,aps,twocolumn]{revtex}

\catcode`\@=11

\def\maketitle2{\par 
\begingroup
\let\cite\@bylinecite
\def\thefootnote{\fnsymbol{footnote}}%
\twocolumn[\@maketitle2\vskip2pc]%
\thispagestyle{plain}\@thanks
\endgroup
\def\thefootnote{\arabic{footnote}}%
\setcounter{footnote}{0}%
\let\maketitle2\relax \let\@maketitle2\relax
\let\@thanks\relax \let\@authoraddress\relax \let\@title\relax
\let\@date\relax \let\thanks\relax \let\@abstract\relax 
\let\@pacs\relax}

\def\abstract#1{\gdef\@abstract{{\par 
\bgroup
\ifdim\prevdepth=-1000pt \prevdepth0pt\fi
\hsize\columnwidth
\dimen0=-\prevdepth \advance\dimen0 by17.5pt \nointerlineskip
\small\vrule width 0pt height\dimen0 \relax}{~~}#1\egroup}}

\def\pacs#1{\gdef\@pacs{{\par 
\bgroup
\hsize\columnwidth \parindent0pt
\ifdim\prevdepth=-1000pt \prevdepth0pt\fi
\dimen0=-\prevdepth \advance\dimen0 by20pt\nointerlineskip
\egroup} PACS numbers:~#1}}

\def\@maketitle2{
\@preprint
\@title
\ifdim\prevdepth=-1000pt \prevdepth0pt\fi
\@authoraddress
\@date
\begin{list}{}{\leftmargin=0.10753\textwidth \rightmargin=\leftmargin
\itemsep=1pc\partopsep=-1pc}
\item\@abstract
\item\@pacs
\end{list}
}

\catcode`\@=12

\begin{document}
\draft
\def\simlt{\stackrel{<}{{}_\sim}}
\def\simgt{\stackrel{>}{{}_\sim}}
\title{Conformal Invariance and Cosmic Background Radiation}    
\author{Ignatios Antoniadis$^1$, Pawel O. Mazur$^2$, and 
Emil Mottola$^3$} 
\preprint{CPTH--S474.1096}
\preprint{LA-UR-96-4026}
\address{$^1$ Centre de Physique Th\'eorique, Ecole 
Polytechnique, F-91128 Palaiseau, France}
\address{$^2$ Dept. of Physics and Astronomy, Univ. of S. 
Carolina, Columbia, SC 29208, USA}
\address{$^3$ Theoretical Division, Los Alamos National 
Laboratory,
MS B285, Los Alamos, NM 87545, USA}
\date{\today}
\abstract
{\small{The spectrum and statistics of the cosmic microwave 
background radiation (CMBR) are investigated under the hypothesis 
that scale invariance 
of the primordial density fluctuations should be promoted to full 
conformal
invariance. As in the theory of critical phenomena, this 
hypothesis leads 
in general to deviations from naive scaling. The spectral index 
of
the two-point function of density fluctuations is given in terms 
of the quantum
trace anomaly and is greater than one, leading to less power at 
large 
distance scales than a strict Harrison-Zel'dovich spectrum. 
Conformal
invariance also implies non-gaussian statistics for the higher 
point
correlations and in particular, it completely determines the 
large angular
dependence of the three-point correlations of the CMBR.}}
\maketitle2
\narrowtext

With the discovery of cosmic microwave background anisotropy 
\cite{cobe},
cosmology has accelerated its transition from a field based 
largely on
speculation to one in which observational data can be brought to 
bear on our
understanding of the universe. As a newly emerging physical 
science, it is
appropriate and may prove fruitful to examine the consequences 
for cosmology of
progress in its more developed sister sciences. 

Scale invariance was first introduced into physics in early 
attempts to
understand the apparently universal behavior observed in 
turbulence and second
order  phase transitions, which are independent of the particular 
dynamical
details of the system. The gradual refinement and 
development
of this
simple idea of universality has led to the theory of critical 
phenomena, which
is remarkable both for its broad applicability and quantitative 
predictive power \cite{RG}.
One of the hallmarks of the modern theory of critical phenomena 
is well-defined
logarithmic deviations from naive scaling relations based on 
engineering
dimensions. A second general feature of the theory is the 
specification of
higher point correlation functions of fluctuations according to 
the requirements
of conformal invariance at the critical point \cite{pol}. 

In cosmology scale invariance was first raised to a level of 
prominence by the
pioneering work of Harrison and Zel'dovich on the spectrum of 
primordial density
fluctuations required to produce the observed large scale 
structure in the
universe \cite{hz}. In inflationary scenarios density 
fluctuations 
with a spectral index $n$ very close to one 
can be generated from quantum fluctuations at a very early stage 
in
the history of the universe at the threshold of its classical 
evolution
\cite{infl}. The most direct probe of these primordial density 
fluctuations is
the cosmic microwave background radiation (CMBR). With the 
observational 
data of the CMBR anisotropy providing confirmation of this 
speculative
foundation, the time now seems ripe to go one step further in 
developing 
the theoretical framework and apply the considerable lessons of 
universality 
in critical phenomena to the universe itself.

In the language of critical phenomena, the observation of 
Harrison and
Zel'dovich
that the primordial density fluctuations should be characterized 
by a spectral
index $n=1$ is equivalent to the statement that the observable 
giving rise to
these fluctuations has engineering or naive scaling dimension 
$\Delta_0 = 2$. Indeed, because the density fluctuations are 
related to
metric fluctuations by Einstein's equations, this naive scaling 
dimension
simply reflects the fact that the relevant coordinate invariant 
measure
of metric fluctuations is the scalar curvature $\delta R \sim 
G\delta\rho$,
which is second order in derivatives of the metric. Hence, the 
fluctuations in
the density perturbations are tied to the scalar curvature and 
the two-point
correlations of both should behave like $|x-y|^{-4}$, or $|k|^1$ 
in Fourier
space, according to simple dimensional analysis.

One of the principal lessons of the modern understanding of 
critical phenomena is that 
naive dimensional
analysis does not fix the transformation properties of 
observables under conformal
transformations at the fixed point, but that instead one must 
expect to find
well-defined logarithmic deviations from naive scaling, 
corresponding to a
(generally non-integer) dimension $\Delta \ne \Delta_0$. The 
deviation from
naive scaling $\Delta - \Delta_0$ is the ``anomalous'' dimension 
of the 
observable due to critical fluctuations which may be quantum or 
statistical in
origin. The requirement of conformal invariance then determines 
the form of the two- and three-point correlation functions of the 
observable in terms of 
its dimension $\Delta$, up to an arbitrary amplitude.

{\em Two-point correlations.} In the case of the two-point 
function of two observables ${\cal O}_{\Delta}$
with dimension $\Delta$, conformal invariance requires
\begin{equation}
\langle{\cal O}_{\Delta} (x_1) {\cal O}_{\Delta} (x_2)\rangle
\sim \vert x_1-x_2 \vert^{-2\Delta}
\end{equation}
at equal times in three dimensional flat spatial coordinates. In
Fourier space this becomes
\begin{equation}
G_2(k) \equiv\langle\tilde{\cal O}_{\Delta} (k) \tilde {\cal 
O}_{\Delta}
(-k)\rangle \sim \vert k \vert^{2\Delta - 3} \,.
\label{G2}
\end{equation}
Thus, we define the spectral index of this observable by
\begin{equation}
n \equiv 2 \Delta - 3\ .
\label{index}
\end{equation}
In the case that the observable is the primordial density 
fluctuation $\delta
\rho$, and in the classical limit where its anomalous dimension 
vanishes, 
$\Delta \rightarrow \Delta_0 =2$, we recover the 
Harrison-Zel'dovich
spectral index of $n=1$.

In order to convert the power spectrum of primordial density 
fluctuations
to the spectrum of fluctuations in the CMBR at large angular 
separations
we follow the standard treatment \cite{SW,peeb}, relating the
temperature deviation to the Newtonian gravitational potential 
$\varphi$
at the last scattering surface, ${\delta T \over T} \sim \delta 
\varphi$,
which is related to the density perturbation by 
\begin{equation}
\nabla^2 \delta\varphi = 4\pi G\, \delta\rho \ .
\label{lap}
\end{equation}
Hence, in Fourier space 
\begin{equation}
{\delta T \over T} \sim {1\over k^2}{\delta\rho\over \rho}\ ,
\end{equation}
and the two-point function of CMBR temperature fluctuations is
determined by the conformal dimension $\Delta$ to be
\begin{eqnarray}
&&C_2(\theta) \equiv \left\langle{\delta T \over T}(\hat r_1)
{\delta T \over T}(\hat  r_2)\right\rangle
\sim \nonumber\\
&&\int d^3 k\left({1\over k^2}\right)^2 G_2(k) e^{i k\cdot 
r_{12}}
\sim \Gamma (2-\Delta) (r_{12}^2)^{2 - \Delta}\ ,
\label{C2}
\end{eqnarray}
where $r_{12} \equiv (\hat r_1 - \hat r_2)r$
is the vector difference between the two positions from which the 
CMBR photons
originate. They are at equal distance $r$ from the observer by 
the 
assumption that
the photons were emitted at the last scattering surface at equal 
cosmic
time. Since $r_{12}^2 = 2 (1- \cos \theta)r^2$, we find then
\begin{equation}
C_2(\theta) \sim \Gamma (2-\Delta) (1-\cos\theta)^{2 - \Delta} 
\end{equation}
for arbitrary scaling dimension $\Delta$.  

The meaning of the pole in the $\Gamma$ function at $\Delta$ 
equal to $2$ is 
best understood by expanding the function $C_2(\theta)$ in
multipole moments,
\begin{equation}
C_2(\theta) = {1\over 4\pi} \sum_{\ell =0}^{\infty} (2\ell + 1)
c_{\ell}^{(2)}(\Delta) P_{\ell} (\cos \theta)\ ,
\label{c2m}
\end{equation}
with
\begin{equation}
c_{\ell}^{(2)}(\Delta) \sim \Gamma(2-\Delta) \sin\left[ \pi 
(2-\Delta)\right]
{\Gamma (\ell + \Delta -2) \over \Gamma (\ell + 4 -\Delta)}\ ,
\end{equation}
which shows that the pole singularity appears only in the $\ell = 
0$ monopole
moment. This singularity occurs due to the integration over the 
whole space
and is just the reflection of the fact that the Laplacian in 
(\ref{lap})
cannot be inverted on constant functions. Since the CMBR 
anisotropy is defined
by removing the isotropic monopole moment (as well as the dipole 
moment, which
receives a substantial contribution from the proper motion of the 
earth with
respect to the CMBR \cite{peeb}), the $\ell =0$ moment does not 
appear in the
sum, and the higher moments of the anisotropic two-point 
correlation function
are well-defined for $\Delta$ near $2$. Normalizing to the 
quadrupole moment
$c_2^{(2)}(\Delta)$, we find
\begin{equation}
c_{\ell}^{(2)}(\Delta) = c_2^{(2)}(\Delta) 
{\Gamma (6 - \Delta) \over \Gamma (\Delta) } 
{\Gamma (\ell + \Delta - 2) \over \Gamma(\ell + 4 - \Delta)}\ ,
\end{equation}
which is a standard result \cite{peeb,be}. Indeed, if $\Delta$ is 
replaced by
$\Delta_0 = 2$ we obtain $\ell (\ell + 1) 
c_{\ell}^{(2)}(\Delta_0) = 6 c_2^{(2)}
(\Delta_0)$, which is the well-known  predicted behavior of the 
lower moments 
($\ell \le 30
$) of the CMBR anisotropy where the Sachs-Wolfe effect should 
dominate.

Up to this point our considerations have been quite general.
In order to say something more definite about the expected 
violations of classical scale invariance we need a specific 
proposal for their physical origin.
This is provided by consideration of the zero-point quantum 
fluctuations 
of massless fields, which give rise to the conformal trace 
anomaly 
${T^{\mu}}_{\mu} \ne 0$. Such a non-zero 
trace 
associated with long-range fields couples to the spin$-0$ or 
conformal part of the metric, and cause it to fluctuate as well. 
These fluctuations of the metric grow logarithmically at distance 
scales
of the order of the horizon and lead to a renormalization group 
flow to 
an infrared stable conformal invariant fixed point of gravity 
\cite{am}.
At this fixed point conformal invariance is restored but there 
are
well-defined deviations of the scaling dimensions of observables 
from
their classical values, calculable in terms of the coefficient
of the original trace anomaly.  

This analysis determines the general form of the conformal 
dimension of an observable with naive dimension $\Delta_0$ to
be \cite{spike}
\begin{equation}
\Delta  =  4{ \sqrt{1 - 2 (4-\Delta_0)/Q^2} - \sqrt{1 - 
8/Q^2}\over
1 - \sqrt{1 - 8/Q^2}}\ ,
\label{dress}
\end{equation}
where $Q^2$ is the coefficient of a certain term (the 
Gauss-Bonnet term) in the
trace anomaly, given by Eqn.~(\ref{cent}) below. 
In the limit $Q^2 \rightarrow \infty$, the effects of
fluctuations in the metric
due to the trace anomaly are suppressed and one recovers the 
classical scaling
dimension $\Delta_0$,
\begin{equation}
\Delta = \Delta_0 + {1\over 2 Q^2} \Delta_0 (4-\Delta_0) + \cdots 
\end{equation}
Hence, consideration of conformal fluctuations of the metric 
generated by
the trace anomaly of massless fields leads necessarily to 
well-defined quantum
corrections to the naive scaling dimensions of observables in 
cosmology.
Moreover, in the analysis of physical observables in the 
conformal sector
of gravity, the operator with the lowest non-trivial scaling 
dimension
corresponds,  in the semi-classical limit, to the scalar 
curvature with
$\Delta_0 = 2$ \cite{spike}. Since the fluctuations which 
dominate at
large distances correspond to observables with lowest 
scaling
dimensions, the conformal factor theory in this limit selects 
precisely 
Harrison's original choice.

With $\Delta_0 = 2$, we find a definite prediction for deviations
from a strict Harrison-Zel'dovich spectrum according to 
Eqns.~(\ref{index})
and (\ref{dress}) in terms of the parameter $Q^2$. The resulting 
spectral index
$n$ is plotted as a function of $Q^2$ in Fig. 1.
It is always greater than $1$
(if $8 \le Q^2 <\infty$), and for large $Q^2$ it behaves as 
$n=1+{4\over Q^2}+\cdots$.
Comparing to the results of the four year COBE DMR data analysis 
of the
power spectrum, $0.9\simlt n_{obs} \simlt 1.5$ \cite{coben}, 
we find that $Q^2_{obs} \simgt 12.4$ from Fig. 1.
\vspace{-2.65cm}
\begin{figure}
\hspace{-0.7cm}
\vspace{-3.5cm}
\centerline{\psfig{figure=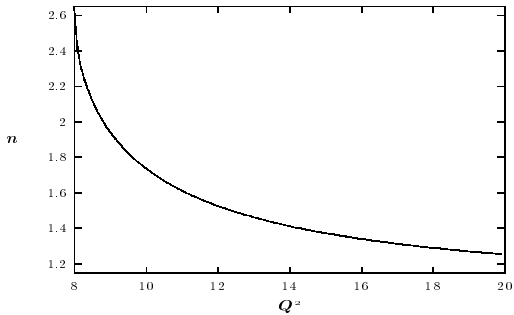,height=13cm,width=12cm}}
\vspace{-2.7cm}
\caption[Figure]{\small{The spectral index $n$ as a function of 
$Q^2$.}}  
\end{figure}
\vspace{-0.2cm} 

{}From the theoretical side, the value of $Q^2$ for free 
conformally invariant
fields is known to be \cite{duff,amm}
\begin{equation}
Q^2 = {1 \over 180}(N_S + {\textstyle{11\over 2}} N_{F} + 62 N_V - 28) + 
Q^2_{grav}\ ,
\label{cent}
\end{equation}
where $N_S, N_{F}, N_V$ are the number of free scalars, Weyl 
fermions and
vector fields and $Q^2_{grav}$ is the contribution of spin-$2$  
gravitons,
which has not yet been determined unambiguously. The $-28$ 
contribution
is that of the conformal or spin-0 part of the metric itself. The 
main
theoretical 
difficulty in
determining $Q^2_{grav}$ is that the Einstein theory is neither 
conformally
invariant nor free, so that a method for evaluating the strong 
infrared effects
of spin-2 gravitons must be found that is insensitive to 
ultraviolet
physics. Such an analysis may be possible by numerical methods on 
the
lattice, which would also provide a nontrivial consistency check 
of the existence of the fixed point with the predicted scaling 
relations \cite{spike}. A purely perturbative computation gives 
$Q^2_{grav}\simeq 7.9$ for the graviton contribution 
\cite{amm}. 
Taking this estimate at face value and including all known 
fields 
of the Standard Model of particle physics (for which
$N_{F}=45$ and $N_V = 12$) we find
\begin{equation}
Q^2_{SM} \simeq 13.2 \qquad {\rm and} \qquad n\simeq 1.45\ ,
\label{standmod}
\end{equation}
which is intriguingly close to the observational bound.

A deviation from the Harrison-Zel'dovich spectrum has 
implications 
for galaxy formation as well. Indeed, a determination of
$Q^2$ close to its
present observational bound together with the COBE quadrupole 
normalization of
the spectrum at large scales implies more power at shorter 
subhorizon scales
where galaxies formed. For example, for the value of the spectral 
index
$n\simeq 1.45$, the power spectrum has an enhancement factor of
$(H_0\times\,20\ {\rm Mpc}/2h)^{-0.45} 
\simeq 4.6$ 
at the $20h^{-1}$ Mpc distance scale, relative to the $n=1$ 
spectrum. 
This would lead to earlier formation of structure at the 
galactic 
and galactic cluster scales than in the case of a primordial 
$n=1$ spectrum. 
However, the form and normalization of the evolved cluster
mass function at these scales is very much model dependent 
and would need to be reanalyzed {\em ab initio} in each dark 
matter 
model to decide if increased power 
in the primordial spectrum of adiabatic density fluctuations can 
be
reconciled with the observations of the matter anisotropy 
on this scale \cite{peeb}. It is noteworthy that the conformal 
invariant 
fixed point 
for gravity predicts a spectral index $n>1$, while most 
suggestions 
for modifying the Harrison-Zel'dovich spectrum such as extended 
or power 
law inflation generally lead to $n\le 1$ \cite{pl}.

{\em Higher point correlations}.
Turning now from the two-point function of CMBR fluctuations to 
higher point
correlators, we find a second characteristic and unambiguous 
prediction of
conformal invariance, namely non-gaussian statistics for the 
CMBR. The first
correlator sensitive to this departure from gaussian statistics 
is the
three-point function of the observable ${\cal O}_{\Delta}$, 
which takes
the form \cite{pol}
\begin{eqnarray}
&&\langle{\cal O}_{\Delta} (x_1) {\cal O}_{\Delta} (x_2) 
{\cal O}_{\Delta} (x_3)\rangle
\sim \nonumber\\
&&\qquad |x_1-x_2|^{-\Delta} |x_2- x_3|^{-\Delta} |x_3 - 
x_1|^{-\Delta}\ ,
\end{eqnarray} 
or in Fourier space,
\begin{eqnarray}
&&G_3 (k_1, k_2) \sim \int d^3 p\ |p|^{\Delta -3}\,  |p + 
k_1|^{\Delta -3}\, 
|p- k_2|^{\Delta -3}\, \sim \nonumber\\
&& \quad\Gamma\left( 3 (1 - {\textstyle{\Delta\over 2}}) \right) \int_0^1 
du\,\int_0^1 dv 
\times \nonumber\\
&&{\left[u(1-u)v\right]^{1-{\Delta\over 2}} 
(1-v)^{-1 + {\Delta\over 2}}\over
\left[u(1-u)k_1^2 + v(1-u)k_2^2 + uv(k_1 + 
k_2)^2\right]^{3(1-{\Delta\over
2})}}\ .
\label{three}
\end{eqnarray}
This three-point function of primordial density fluctuations 
gives rise to
three-point correlations in the CMBR by reasoning precisely 
analogous as that
leading from Eqns.~(\ref{G2}) to (\ref{C2}). That is,
\begin{eqnarray}
&&C_3(\theta_{12}, \theta_{23}, \theta_{31}) 
\equiv \left\langle{\delta T \over T}(\hat r_1)
{\delta T \over T}(\hat  r_2){\delta T \over T}(\hat  
r_3)\right\rangle
\sim \nonumber\\
&&\int {d^3 k_1\,d^3k_2 \over
k_1^2 k_2^2 (k_1 + k_2)^2} G_3 (k_1, k_2) e^{i k_1\cdot r_{13}} 
e^{ik_2\cdot
r_{23}}\ ,
\label{C3}
\end{eqnarray}
where $r_{ij}\equiv ({\hat r}_i-{\hat r_j})r$ and 
$r_{ij}^2=2(1-\cos\theta_{ij})
r^2$.

{}From the above expressions, it is easy to extract the global 
scaling of the
three-point function in the infrared:
\begin{eqnarray}
G_3(\lambda k_1,\lambda k_2) &\sim& \lambda^{3(\Delta -2)} 
G_3(k_1,k_2)\ ,
\nonumber\\
C_3 &\sim& r^{3(2-\Delta)}\ .
\label{gloscal}
\end{eqnarray}
In the general case of three different angles, the expression for 
the 
three-point correlation function (\ref{C3}) is quite complicated, 
though
it can be rewritten in parametric form analogous to (\ref{three})
to facilitate numerical evaluation, if desired. 
An estimate of its angular dependence in the limit 
$\Delta\to 2$ can be obtained by replacing the slowly varying 
$G_3(k_1,k_2)$ by a
constant. Then (\ref{C3}) can be evaluated by expanding
in terms of spherical harmonics:
\begin{eqnarray}
&&C_3(\theta_{ij})\sim
\sum_{l_i,m_i}{{\cal K}^*_{l_1m_1l_2m_2l_3m_3}\over
(2l_1+1)(2l_2+1)(2l_3+1)} 
\left( {1\over l_1+l_2+l_3}+ \right.\nonumber\\
&&\quad\left.{1\over l_1+l_2+l_3+3}\right) Y_{l_1m_1}({\hat r}_1) 
Y_{l_2m_2}({\hat r}_2)
Y_{l_3m_3}({\hat r}_3)\ ,
\end{eqnarray}
where
${\cal K}_{l_1m_1l_2m_2l_3m_3}\equiv\int d\Omega 
Y_{l_1m_1}(\Omega)
Y_{l_2m_2}(\Omega)  Y_{l_3m_3}(\Omega)$.

In the special case of equal angles $\theta_{ij}=\theta$
\cite{cobe3}, it follows from (\ref{gloscal}) that the
three-point correlator is 
\begin{equation}
C_3(\theta)\sim (1-\cos\theta)^{{3\over 2}(2-\Delta)}\ .
\end{equation}
Expanding the function $C_3(\theta)$ in multiple moments as in 
Eqn.~(\ref{c2m})
with coefficients $c_{\ell}^{(3)}$, and normalizing to the 
quadrupole moment,
we find
\begin{equation}
c_{\ell}^{(3)}(\Delta) =c_{2}^{(3)}(\Delta)
{\Gamma (4+{3\over 2}(2-\Delta))\over\Gamma (2-{3\over 
2}(2-\Delta))}
{\Gamma (\ell-{3\over 2}(2-\Delta))\over\Gamma(\ell+2+{3\over 
2}(2-\Delta))}\ .
\label{cl3}
\end{equation}
In the limit $\Delta=2$, we obtain 
$\ell(\ell+1)c_{\ell}^{(3)}=6c_2^{(3)}$,
which is the same result as for the moments $c_{\ell}^{(2)}$ of 
the two-point 
correlator but with a different quadrupole amplitude.

The value of this quadrupole normalization $c_2^{(3)}(\Delta)$ 
cannot 
be determined by conformal symmetry considerations alone. A naive
comparison with the two-point function which has a 
small amplitude
of the order of $10^{-6}$ leads to a rough estimate of 
$c_2^{(3)}\sim{\cal
O}(10^{-9})$, which would make it very difficult to detect 
\cite{cobe3}.
However, if the conformal invariance hypothesis is correct, 
then these non-gaussian correlations must exist at some level, in  
distinction to the simplest inflationary scenarios. 
Their amplitude is model dependent and possibly much 
larger
than the above naive estimate. 
The detection of such non-gaussian correlations at any level is 
therefore 
an
important test for the hypothesis of conformal invariance.

{}For higher point correlations, conformal invariance does not 
determine
the total
angular dependence. Already the four-point function takes the 
form,
\begin{equation}
\langle{\cal O}_{\Delta} (x_1) {\cal O}_{\Delta} (x_2) 
{\cal O}_{\Delta} (x_3) {\cal O}_{\Delta} (x_4)\rangle
\sim { A_4
\over {\prod_{i<j} r_{ij}^{2\Delta/3}} }\ ,
\end{equation}
where the amplitude $A_4$ is an arbitrary function of the two 
cross-ratios, 
$r_{13}^2 r_{24}^2/r_{12}^2 r_{34}^2$ and 
$r_{14}^2 r_{23}^2/r_{12}^2 r_{34}^2$.
Analogous expressions hold for higher $p$-point functions. 
However in the
equilateral case $\theta_{ij}=\theta$, the coefficient amplitudes 
$A_p$ become 
constants and the
angular dependence is again completely determined. The result is
\begin{equation}
C_p(\theta)\sim (1-\cos\theta)^{{p\over 2}(2-\Delta)}\ ,
\end{equation}
and the expansion in multiple moments yields coefficients 
$c_{\ell}^{(p)}$
of the
same form as in Eqn.~(\ref{cl3}) with $3/2$ replaced by $p/2$.
In the limit $\Delta =2$, we obtain the universal 
$\ell$-dependence
$\ell(\ell+1)c_{\ell}^{(p)}=6c_2^{(p)}$.

In summary, the conformal invariance hypothesis applied to
the primordial density fluctuations predicts 
deviations from the Harrison-Zel'dovich spectrum, which should
be imprinted on the CMBR anisotropy. A particular realization
of this hypothesis is provided by the metric fluctuations induced
by the known trace anomaly of massless matter fields which
gives rise to fixed point with a spectral index $n> 1$. A second
general consequence of conformal invariance is non-gaussian 
higher
point correlations in the statistics of the CMBR. If either
of these effects is detected it would be an important clue to
the mechanism of the origin of primordial density fluctuations 
and the formation of structure in the universe. 

P.O.M. and E.M. thank the CPhT of the Ecole Polytechnique for its
hospitality. Research supported in part by EEC contract 
CHRX-CT93-0340.


\end{document}